\documentclass[12pt]{iopart}
\usepackage{iopams}  
\usepackage{graphicx}
\usepackage{color}
\begin{document}

\title[Reconnection in MAST and the corona]{Two-fluid and magnetohydrodynamic modelling of magnetic reconnection in the MAST spherical tokamak and the solar corona
}

\author{P K Browning$^1$, S Cardnell$^1$, M Evans$^1$, F Arese Lucini$^1, ^2 $, V S Lukin$^3$, K G McClements$^4$ and A Stanier$^1,^5$}

\address{$^1$Jodrell Bank Centre for Astrophysics, University of Manchester, UK,

$^2$ City University New York Graduate Centre, New York USA,

$^3$ National Science Foundation, Arlington, Virginia, USA (formerly of US Naval Research Laboratory),

$^4$ CCFE, Culham Science Centre, Abingdon, UK,

$^5$ Los Alamos National Laboratory, Los Alamos, USA}
\ead{p.browning@manchester.ac.uk}

\section*{abstract}
{Twisted magnetic flux ropes are ubiquitous in laboratory and astrophysical plasmas, and the merging of such flux ropes through magnetic reconnection is an important mechanism for restructuring magnetic fields and releasing free magnetic energy. The merging-compression scenario is one possible  start-up scheme for spherical tokamaks, which has been used on the Mega Amp Spherical Tokamak (MAST). Two current-carrying plasma rings – or flux ropes – approach each due to mutual attraction, forming a current sheet and subsequently merge through magnetic reconnection into a single plasma torus, with substantial plasma heating. Two dimensional resistive and Hall MHD simulations of this process are reported.  A model of the merging based on helicity-conserving relaxation to a minimum energy state is also presented, extending previous work to tight-aspect-ratio toroidal geometry. This model leads to a prediction of the final state of the merging, in good agreement with simulations and experiment, as well as the average temperature rise. A relaxation model of reconnection between two or more flux ropes in the solar corona is also described, allowing for different senses of twist, and the implications for heating of the solar corona are discussed.}

\maketitle

\section{Introduction}

Magnetic reconnection is a process for re-structuring magnetic fields and rapidly converting magnetic energy to thermal energy, with important consequences in many laboratory, space and  astrophysical plasmas \cite{priest00, biskamp00,zweibel09}.  Merging of twisted bundles of magnetic field lines, known as ''flux-ropes'' \cite{lukin14, ryutova15}, through  reconnection is widespread, and  has been investigated in several purpose-built laboratory experiments e.g. \cite{ono93,brown99, furno03, yamada07, gekelman12, tripathi13}. 

The MAST spherical tokamak  \cite{lloyd11} - as well as demonstrating the potential for fusion energy generation of  the spherical tokamak concept,  and  providing insight into many  physical processes relevant to conventional tokamaks - has incidentally provided a valuable testbed for reconnection studies. In the merging-compression start-up scheme  (one of several alternative plasma start-up methods), two flux-ropes with parallel toroidal current move together and then merge, creating a single plasma  torus with closed magnetic flux surfaces \cite{ono12}. This provides start-up and current drive without using  a central solenoid, which could be attractive for future fusion devices,  as well as allowing an experimental study, with good diagnostics, of reconnection and flux-rope merging \cite{ono12, stanier13} in a parameter regime more  closely resembling the solar corona and other astrophysical plasmas \cite{browning14} than other experiments.  
        
The dissipation of stored magnetic energy through  reconnection \cite{longcope15} - which is  the primary source of energy release in solar flares - provides a strong candidate for  resolving the mystery of how  solar coronal plasma is heated to  temperatures of over 10$^6$K 
\cite{demoortel15, klimchuk15}. Such reconnection may occur through merging of twisted magnetic flux ropes \cite{gold60}. In the flux-tube tectonics scenario \cite{priest02}, a coronal loop may contain multiple twisted threads as its field lines are rooted in several discrete photospheric flux sources. Adjacent twisted flux tubes may also be created through photospheric motions with multiple vortices within a single flux source. 

In the next section, we summarise recent results from single-fluid MHD and Hall-MHD simulations of flux-rope merging in MAST,  and discuss the heating of ions and electrons through reconnection. In Section \ref{relaxed_model}, we outline how relaxation theory provides a useful tool for calculating the final state and the energy dissipated, and present a new model of relaxation in a tight-aspect-ratio geometry. Implications for heating of the solar corona, mainly based on relaxation theory, are presented in Section \ref{corona}.

\section{Overview of resistive MHD and Hall-MHD simulations}
\label{simulations}
MAST \cite{lloyd11} is a tight-aspect-ratio toroidal device, with typical major and 
minor radii $R = 0.95\,$m, $a = 0.60\,$m, plasma current $I_p = 400 - 900\,$kA, toroidal field $B_{\rm T} = 0.40 - 0.58\,$T  at $R= 0.7\,$m, and peak electron density and temperature $n_{e0} \simeq 3 \times 10^{19}\,$m$^{-3}$ and $T_{e0} \simeq 1\,$keV. The merging-compression scheme is initiated by the production of  two toroidal flux-ropes with parallel toroidal currents  around the in-vessel P3 poloidal field coils (see Figure 1). As the current in the coils is decreased, the attraction of the ``like'' induced plasma currents in the two flux-ropes causes  them to separate from the coils and move together, subsequently merging through magnetic reconnection into a single flux-rope.  Recent experiments using merging-compression have produced  spherical tokamak plasmas with  currents of up to 0.5 MA. The merging is associated with rapid heating (presumably as a result of reconnection), indicated by measured electron temperatures of up to 1 keV  and ion temperatures up to 1.2 keV \cite{ono12,tanabe15}.

\begin{figure}
\label{mast_cartoon}
\includegraphics[scale=0.8]{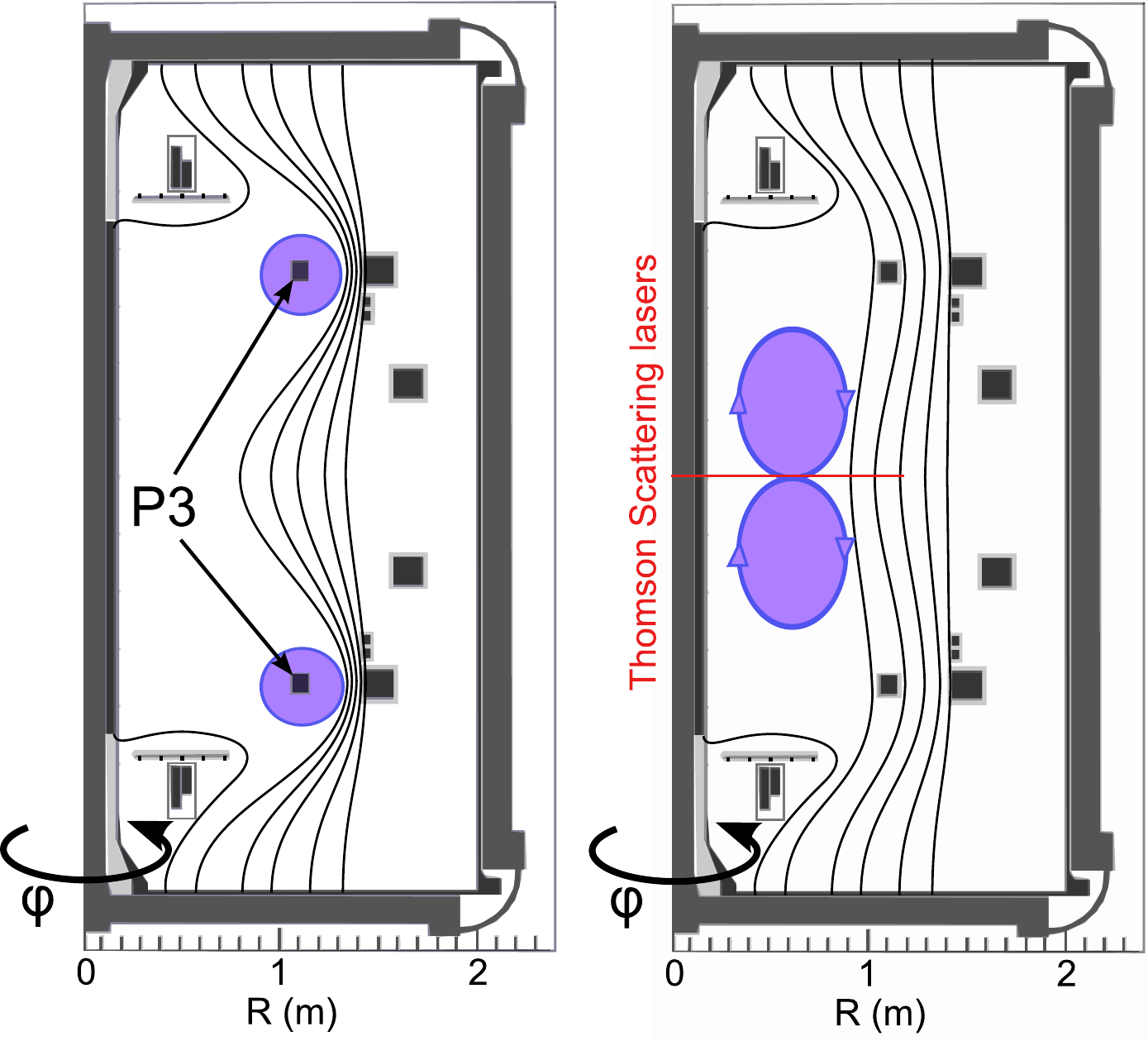}
\caption{Cartoon of merging-compression formation in MAST \cite{meyer13, stanier13}, with flux ropes shown in purple, forming around the P3 poloidal field coils (left panel) and on the point of merging (right panel).}
\end{figure}

Recently,  fully nonlinear and compressible 2D simulations  have been performed of flux-rope merging in MAST using  both single-fluid MHD and Hall-MHD \cite{stanier13}. The Hall terms  in Ohm's Law are considered because, for typical MAST parameters, the ion skin-depth is a macroscopic length scale (normalised ion skin depth $d_i$ = 0.145) and is much larger than the width of the Sweet-Parker current sheet predicted by resistive MHD. 

The dimensionless Hall-MHD equations  \cite{biskamp00} used for the simualations are as follows:  
\begin{equation}\label{mass}\partial_t n + {\bf \nabla} \cdot (n{\bf v}_i) = 0\end{equation}
\begin{equation}\label{mom}\partial_t (n {\bf v}_i) + {\bf \nabla} \cdot (n {\bf v}_i {\bf v}_i + p{\sl I} + {\bf \pi}_i) = {\bf j} \times {\bf B}\end{equation}
\begin{equation}\label{ohm}{\bf E} = -{\bf v}_e\times {\bf B} - \frac{d_i}{n}{\bf \nabla}p_e + \eta {\bf j} - \eta_H \nabla^2 {\bf j}\end{equation}
\begin{equation}\label{faraday}\partial_t {\bf B} = - {\bf \nabla} \times {\bf E}\end{equation}
\begin{equation}\label{pressure}(\gamma - 1)^{-1}\left[\partial_t p + {\bf v}_i \cdot {\bf \nabla}p + \gamma p{\bf \nabla}\cdot {\bf v}_i\right] = \eta j^2 + \eta_H ({\bf \nabla}{\bf j})^2 - {\bf \pi}_i:{\bf \nabla}{\bf v}_i - {\bf \nabla} \cdot {\bf q}\end{equation}
Normalisation is with respect to typical values in  MAST start-up plasmas prior to merging: density $n_0 = 5 \times 10^{18}$ m$^{-3}$, length-scale $L_0 = 1$ m, magnetic field $B_0 = 0.5$ T, and (equal) ion and electron temperatures $T_0= 10$ eV; velocities are normalised by the Alfv\'en speed $v_0 = B_0 (\mu_0 n_0 m_i)^{-1/2}$. Here,  $n$ is the density, ${\bf j} = {\bf \nabla}\times {\bf B}$ the current density, ${\bf v}_i$ the ion  velocity (with electron velocity ${\bf v}_e = {\bf v}_i - d_i {\bf j}/n$), ${\bf B}$ the magnetic field, $p=p_i+p_e$ the total (sum of the ion and electron) thermal pressure and  ${\bf E}$ the electric field. The ion stress tensor is ${\bf \pi}_i = - \nu_i({\bf \nabla}{\bf v}_i + {\bf \nabla}{\bf v}_i^T)$, and the heat-flux vector ${\bf q}$ has anisotropic form ${\bf q} = -\kappa^\parallel_e {\bf \nabla}_\parallel T - \kappa^\perp_i {\bf \nabla}_\perp T$ where ${\bf \nabla}_\parallel = {\bf \hat{b}}({\bf \hat{b}}\cdot {\bf \nabla})$. The coefficients are (all normalised):  ion skin-depth, $d_i$,  parallel resistivity  $\eta$, parallel ion viscosity  $\nu_i$ and  parallel electron, $\kappa_e^\parallel$, and perpendicular ion, $\kappa_i^\perp$, heat conductivities; here, we set the normalised values to be  $\kappa_e^\parallel=10^{-1}$ and  $\kappa_i^\perp =10^{-7}$.  The final term on the right-hand side of equation~(\ref{ohm}) is hyper-resistive diffusion \cite{biskamp00}, used only the Hall-MHD simulations, which represents anomalous electron viscosity and sets a dissipation scale for Whistler and kinetic Alfven waves, with normalised hyper-resistivity $\eta_H$.  Normalised Braginskii values of resistivity and parallel viscosity in MAST start-up conditions  are $\eta  = 10^{-5}$ and $\nu_i  = 10^{-3}$; but values of  $\eta$, $\nu_i$ and $\eta_H$ are  varied in order to study the scaling effects of collisions. 

Simulations were performed in: cartesian geometry, with invariance in the ``toroidal'' direction, representing an infinite aspect ratio system; and an axisymmetric tight aspect ratio toroidal system, invariant in the toroidal $\phi$ direction. The initial configuration was chosen to represent the instant when the current rings have detached from in-vessel coils,  resulting in two localised cylinders/tori  of toroidal current. The toroidal field within these flux ropes was calculated to ensure local force balance, but a non-zero attractive force between the ``like''  parallel currents  caused the rings to move towards each other and eventually reconnect.  In the tight aspect ratio case, a confining vertical field was also  incorporated. 

\begin{figure}
\center
\includegraphics[width=1.0\textwidth]{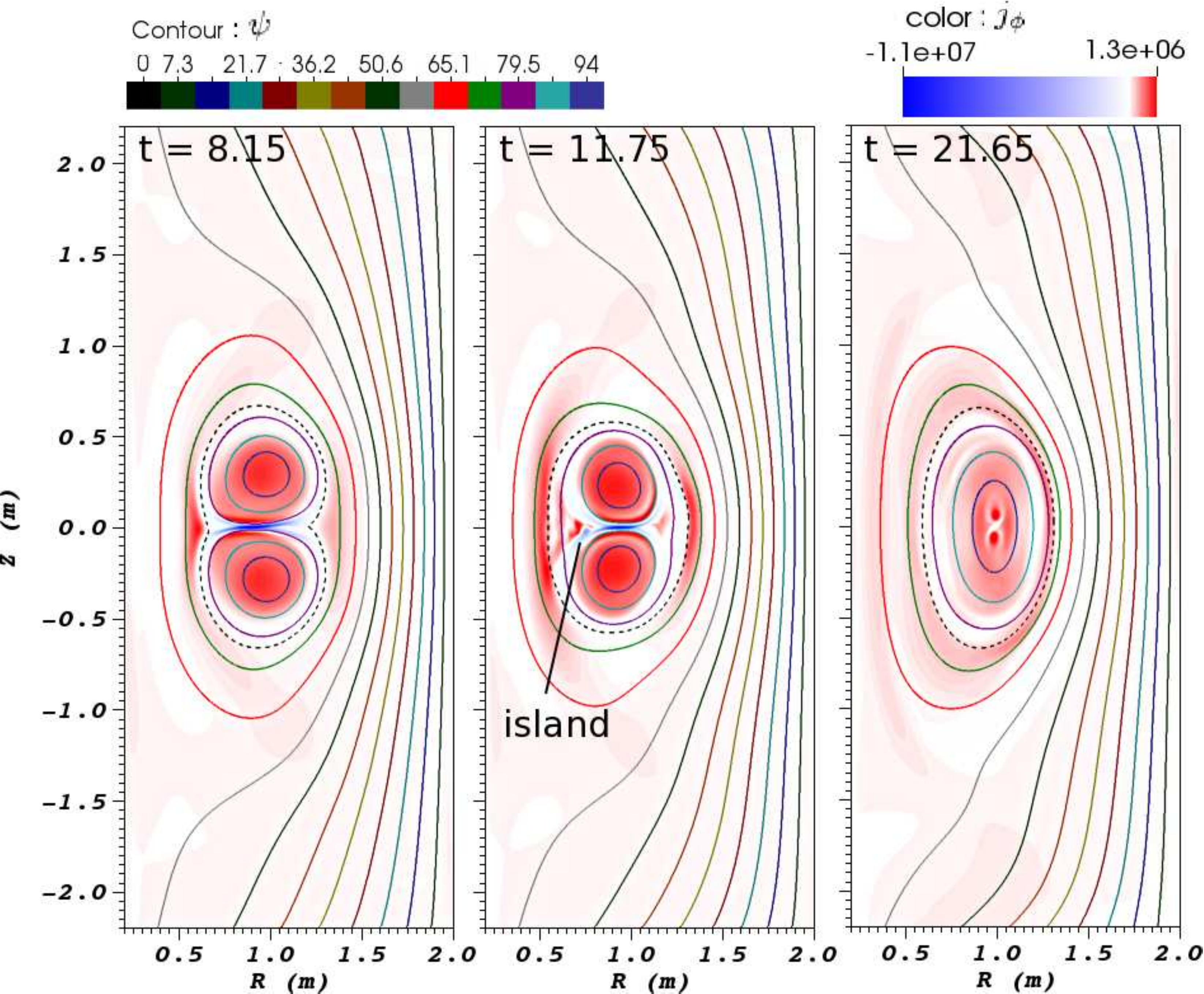}

\caption{Snapshots showing poloidal field (contours) and  toroidal current density (colour scale) for Hall MHD simulation  of plasma merging in MAST. Toroidal geometry was used, with $\eta_{H}= 10^{-8}$. }
\label{Hall_merging}
\end{figure}

Results from a Hall-MHD simulation in toroidal geometry are shown in Figure 2. The key results of these simulations are as follows:
\begin{itemize}
\item The flux-ropes eventually merge into a single flux-rope. For the simulations in toroidal geometry, the final state is a single torus with nested flux-surfaces  and magnetic field profiles resembling those found in MAST plasmas.

\item Oscillations  (``sloshing'') in the reconnection rate at low resistivity, resulting from magnetic pressure pile-up.

\item A much faster reconnection rate in Hall MHD than in resistive MHD.

\item The qualitative nature of the reconnection in Hall-MHD simulations depends on the ratio of the collisional current sheet width - determined by the hyper-resistivity - to the ion-sound Larmor radius. At higher collisionality there is a broad current sheet, whilst at intermediate values secondary tearing creates small islands of poloidal flux. At low collisionality, the outflow separatrices open and fast reconnection is attained.

\item Toroidal simulations in Hall-MHD  show the formation of a double-peaked structure in the radial density profile which closely resembles experimentally-measured profiles.

\end{itemize}
Finally, we note that preliminary work has been undertaken to predict the evolution of ion and electron temperatures, using separate equations for ion and electron temperature evolution incorporating, respectively, ion viscous heating and resistive and hyper-resistive electron heating \cite{stanier14}. Plots of the evolution of both temperature profiles are shown in Figure 3. Ions are mainly heated by collisional viscous heating in the reconnection outflow jets, giving a  temperature profile which is double-peaked radially, consistent with experiment \cite{tanabe15}. The simulations predict   ion peak temperatures of the order of 1 keV, comparable with the observations of Ono et al \cite{ono12}, but higher than reported by Tanabe et al  \cite{tanabe15};  the latter  experiments were undertaken at low coil currents. The predicted electron temperature profiles are dominated by hyper-resistive diffusion which is large in the presence of strong gradients of the current density, and so is not cospatial with the current density; this is also suggested by the experimental profiles. The experimental observation that the electron temperature is sometimes centrally-peaked, and sometimes hollow, may be linked to the ``stochastic'' appearance and ejection of plasmoids in the simulations. Further work is in progress on simulating the evolution of electron and ion temperature profiles.

\begin{figure}
\center
\includegraphics[width=1.0\textwidth]{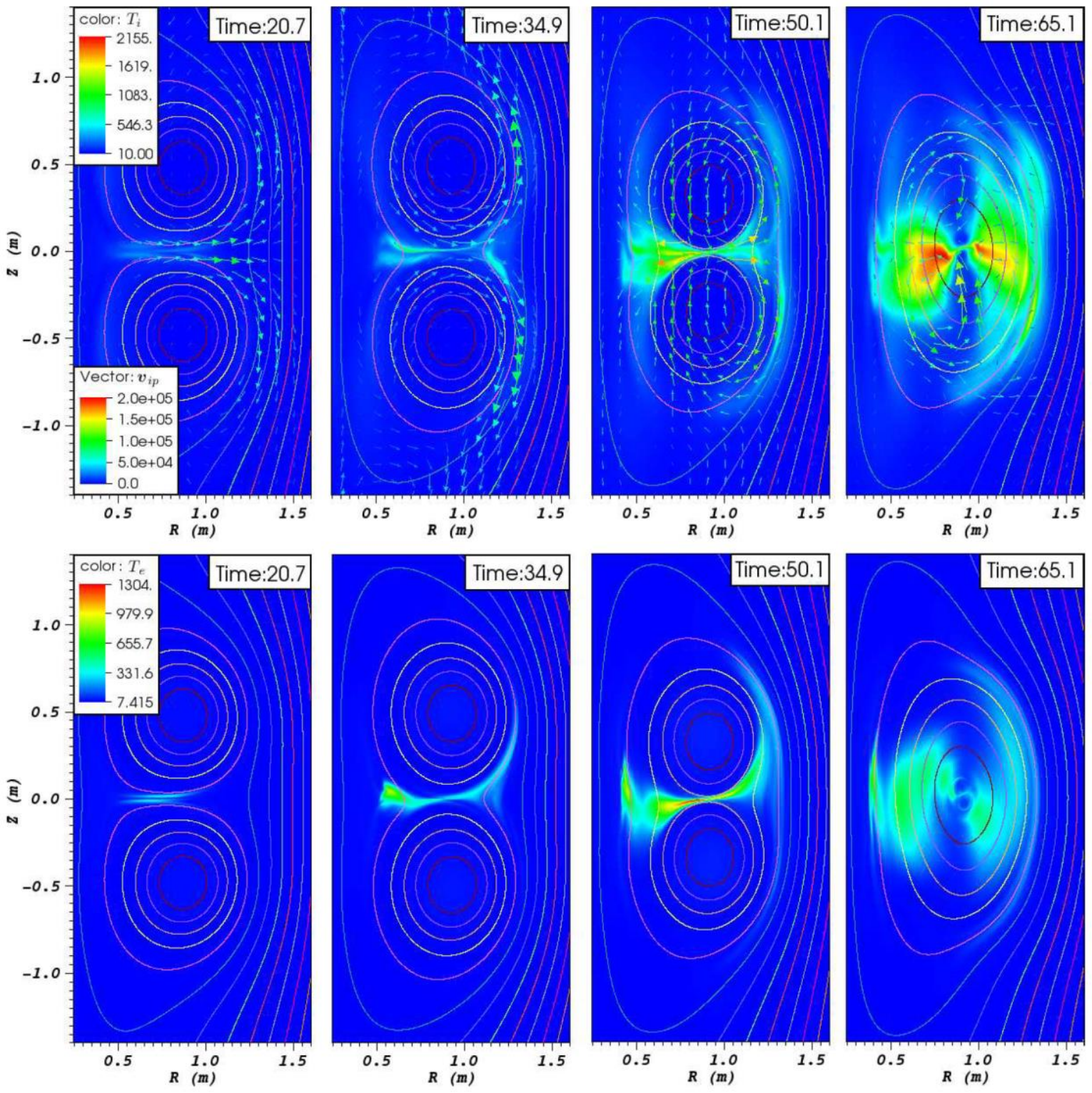}

\caption{Ion (top) and electron (bottom) temperature profiles, in eV, at four successive times, for Hall MHD simulation  of plasma merging in MAST. The top plot also shows in-plane ion velocity vectors. Toroidal geometry was used, with $\eta_{H}= 10^{-8}$. }
\label{temperature}
\end{figure}

\section{Relaxation and self-organisation in merging flux ropes in spherical tokamaks}
\label{relaxed_model}
Relaxation theory, first proposed by Taylor to explain the field configurations in RFP experiments \cite{taylor74}, has provided a powerful tool for explaining phenomena in various magnetic confinement devices \cite{taylor86} as well as astrophysical plasmas. It is hypothesised that a disrupted magnetic field in a highly-conducting plasma relaxes to a state of minimum magnetic energy whilst conserving global magnetic helicity $K=\int_V {\bf A} \cdot {\bf B}dV$. The relaxed state is a linear or constant-$\mu$  field:

\begin{equation}
   \label{constmu}
      \nabla\times{\bf B} = \mu{\bf B},
   \end{equation}
where $\mu = \mu_{0}{\bf B} \cdot {\bf j}/B^{2}$  is spatially constant. Note that the relaxed state is a special case of a force-free, satisfying $\bf{j} \times \bf{B} = 0$.

It was recently proposed \cite{browning14}  that the merging of flux ropes  in MAST could  be modelled as a helicity-conserving  relaxation to a minimum energy state.  The model assumes two initial force-free flux ropes separated by a current sheet (a  discontinuity in the poloidal field); these merge into a lower energy state, consisting of a single flux rope within the same volume described by (\ref{constmu}). In order to permit analytical solutions, it is   assumed that each  initial flux rope has a linear
force-free field  described by (\ref{constmu}). The  initial field  has free energy associated with the  current sheet between the two flux ropes (this is a delta function layer of reversed current, or negative $\mu$), and hence can release energy through relaxation to a fully constant-$\mu$ state; it is, however, a force-free equilibrium.  This represents the field at the moment  at which the flux ropes have been brought together by the attractive force, but have not yet commenced reconnection.  Browning et al \cite{browning14} developed this model assuming infinite aspect-ratio i.e. the flux ropes are straight cylinders; a Cartesian coordinate system is used with all quantities independent of the axial coordinate $z$.
 
Since MAST, and several other laboratory experiments  involving flux rope merging, have very low aspect-ratio, it is
interesting to extend  the  relaxation model of \cite{browning14} to finite aspect-ratio. We therefore model relaxation of two toroidal flux ropes within a cylindrical "can"  of rectangular cross-section (hence still allowing analytical solutions). The spherical tokamak is assumed to be an axisymmetric configuration contained within the volume  $0 \leq Z \leq l; a \le R \leq b$ in cylindrical coordinates $(R, \phi, Z)$. Lengths are normalised with respect to the width of the can, $b-a$ (the minor radius is $(b-a)/2$, and the aspect ratio is $(b+a)/(b-a)$. Similarly to  the infinite-aspect-ratio model, the initial state is given by two constant-$\mu$ toroidal flux ropes separated by an annular current sheet at the midplane  $Z = 0.5 l$, where we take $l =2$ (in dimensionless units), giving each flux-rope a square boundary similar to the earlier infinite-aspect ratio model \cite{browning14}. 

The fields (for both initial and final states) are expressed in terms of  a flux function $\psi$ as  ${\bf B} = (1/R)(-\partial\psi/\partial Z, \mu\psi, \partial\psi/\partial R)$ and equation (\ref{constmu}) leads to the Grad Shafranov equation
 \begin{equation}
\label{gs}
{\partial^2\psi\over \partial R^2} - {1\over R}{\partial\psi\over \partial R} + {\partial^2\psi\over \partial Z^2}+\mu^2\psi=0.
\end{equation}
The solution closely follows \cite{browning14}, converted from cartesian to cylindrical coordinates. The boundary condition is $\psi= \psi_b$ = constant on the boundary, where $\psi_b$ (effectively a normalisation constant for the fields) must be non-zero.  (For $\psi_b =0$, the only non-trivial solutions to (\ref{gs}) are spheromak-like eigenfunction solutions for discrete eigenvalues of $\mu$ \cite{browning93}). Following \cite{jensen84} and \cite{browning93}, the solution within a annular shell with height $L$ is expressed as a sum of the eigenfunctions
 \begin{equation}
\label{psi}
\psi=\psi_b\left[1 +  \sum\limits_{m,n} a_{nm}Rr_{m}(R)\sin\left(\frac{n\pi Z}{L}\right)\right], 
\end{equation}
where the radial eigenfunctions are given in terms of Bessel functions $J_{1}, Y_{1}$ as
\begin{equation}
\label{Rm}
r_{m}(x)=Y_{1}(\beta_{m}b)J_{1}(\beta_{m}x)-Y_{1}(\beta_{m}x)J_{1}(\beta_{m}b),
\end{equation}
and $\beta_{m}$ is the {\it m}th zero of the function $r(R)=Y_{1}(bR)J_{1}(aR)- Y_{1}(aR)J_{1}(bR)$.
Note that equation (\ref{psi}) automatically satisfies the required boundary conditions. We  normalise so that the toroidal flux  $\Phi_t \equiv \int\limits_{A}B_z dS =  \int\limits{_0^L} \int\limits{_a^b} \mu \psi dR dZ = 1$ (thereby ensuring conservation of flux during the relaxation process described below): hence $\psi_b(\mu;a,b)$ depends on $\mu$.

Substitution into equation (\ref{gs}) gives
 \begin{equation}
\label{sum}
\sum\limits_{m,n}\left[(-\gamma_{mn}^2+\mu^2)Ra_{nm}r_{m}\left(R/b\right)\sin\left(n\pi Z/b\right)\right]=-\mu^2,
\end{equation}
where
\begin{equation}
\gamma_{mn}^2 = (\beta_{m})^2 + (n\pi/L)^2.
\end{equation}
Multiplying by an  eigenfunction  $r_{m}(R/b)\sin(n\pi Z/L)$ and integrating over $R$ and $Z$, exploiting the orthogonality of these functions, then yields:
\begin{eqnarray}
\label{coeff}
a_{mn}=\frac{\mu^2}{\gamma_{mn}^2-\mu^2} \frac{4}{nb} \frac{1-j_{m}b/a}{1-j_{m}^2}, \qquad\mbox{for $m,n$ odd;}\\
a_{mn}=0 ,  \qquad\mbox{for m or n even;}
\end{eqnarray}
where 
\begin{equation}
j_{m}=J_{1}(\beta_{m}b)/J_{1}(\beta_{m}a).
\end{equation}
The fields from equations (\ref{psi}, \ref{coeff}) are used both to represent the two individual flux ropes in the initial state (with $L = l/2 = b-a $) and the single relaxed flux-rope (with $L = l = 2(b-a)$).  The final value of $\mu$, $\mu_f$, is determined from the initial value in each flux-rope, $\mu_{i}$, by the constraint that both helicity, $K$,and total toroidal flux are conserved (where the dimensionless total toroidal flux initially is 2, because of the two flux-ropes). A root-finding process is used to find $\mu_f$ so as to conserve $K/\Phi_t^2$. The helicity is calculated requiring the toroidal component of the vector potential $A_{\phi}$ to vanish on the boundary so that the gauge-correction term arising from  the flux through the torus \cite{bevir85} vanishes. It can be shown that this is related to the total magnetic energy, $W$, by
\begin{equation}
\label{KW}
K=2\mu_{0}W/\mu-2\pi\psi_{b}\Phi_{t},
\end{equation}
where the dimensionless energy  is obtained by integration of the fields, after use of Fourier identities for the summation over $z$ and considerable algebra following \cite{browning93}, as
\begin{equation}
\label{energy}
W = \frac{\psi_{b}^{2}}{\mu_{0}}\ln\left[\frac{b}{a} + \mu^{2}\sum\limits_{odd m}\frac{2(1-j_{m}b/a)^{2}}{k_{m}^2\beta_{m}b(1-j_{m})^{2}}(P_{m}(\beta_{m}^2+\mu^{2})+k_{m}^2Q_{m}+S_{m})\right], 
\end{equation}
where 
\begin{eqnarray}
k_{m}^{2}=\mu^2-\beta_{m}^2, \\
P_{m} = 1/(\cos(k_{m}L)+1) - 3{\rm tan}(k_{m}L/2)/(k_{m}L)+1,\\
Q_{m}= 2{\rm tan}(k_{m}L/2)/(k_{m}L) -1, \\
S_{m} =(k_{m}^2-k_{m}\sin(k_{m}L)/L)/(\cos(k_{m}L)+1),
\end{eqnarray}
and $\psi_b(\mu,a,b)$ is given by the normalisation $\Phi_t=1$ (and $\mu_{0}=1$ in dimensionless units). 

\begin{figure}
\label{tightaspectratio_fields}
\center
\includegraphics[width=0.3\textwidth]{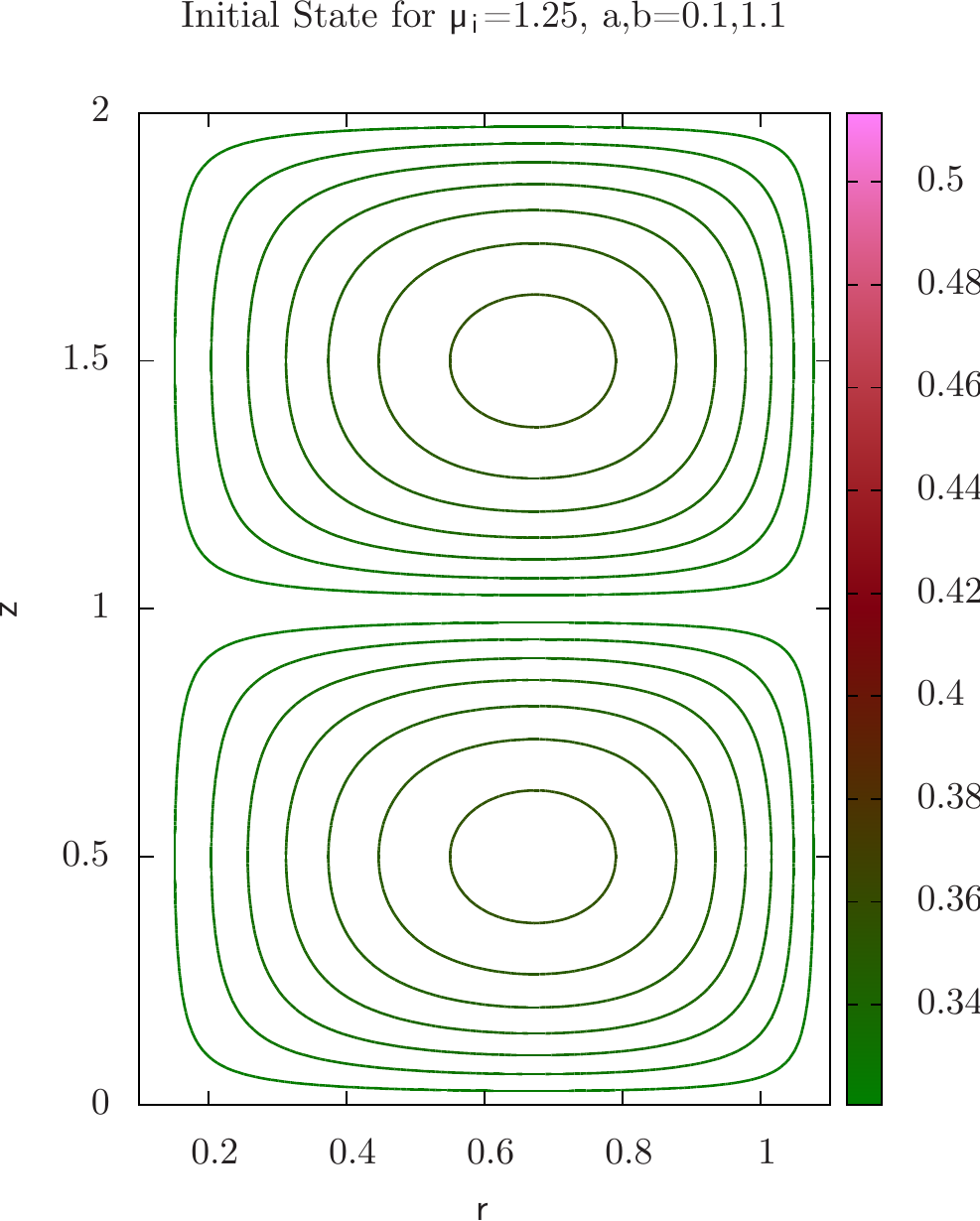} 
\includegraphics[width=0.3\textwidth]{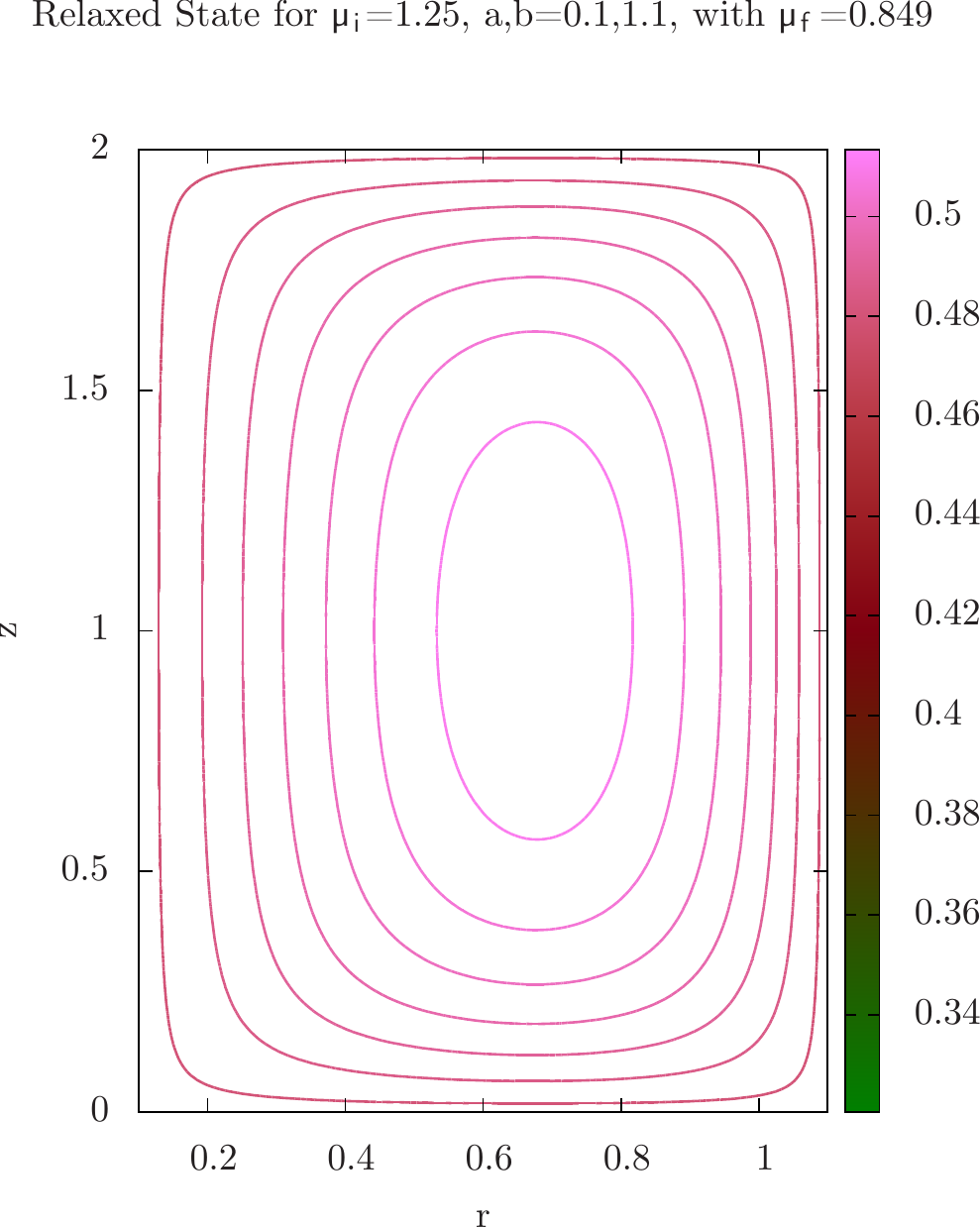}
\caption{Poloidal field lines for initial (left) and final (right)
states for the relaxation model of merging flux ropes in a tight-aspect-ratio configuration with a =0.1, b = 1.1. The initial field (left) has $\mu_{i} = 1.25$. The final field (right) has $\mu_{f}  = 0.849$. In both cases, there is also a toroidal field ($B_Z=\mu\psi$), so that closed poloidal flux contours correspond to twisted magnetic flux ropes.}
\end{figure}

Typical  initial and final field states in a tight-aspect ratio geometry are shown in Figure 4. The energy released during relaxation is calculated as $\Delta W=W_i-W_f$ where the initial energy $W_i=2W(\mu_i;a,b,b-a)$ and the final energy is $W_f=4W(\mu_f;a,b,2(b-a))$ (the factors 2 and 4 are to account for, respectively, two initial flux ropes each with $\Phi_{t}=1$, and a single final  flux rope with $\Phi_{t}= 2$. In Figure 5, the final value of $\mu$ and energy release $\Delta W$ are plotted against the initial value, for both a tight-aspect ratio configuration similar to  MAST and a very large-aspect ratio case. Note that the latter graph very closely resembles the infinite-aspect ratio results \cite{browning14}. Indeed, it can be shown for an aspect-ratio larger than about 10, $\mu$ and $\Delta W$ are almost identical to the infinite-aspect ratio values. However, at lower aspect-ratios,  $\Delta W$  significantly decreases; so that the previous infinite-aspect ratio model over-estimates the energy release for MAST geometry by a factor of around 2.

\begin{figure}
\center
\includegraphics[width=0.7\textwidth]{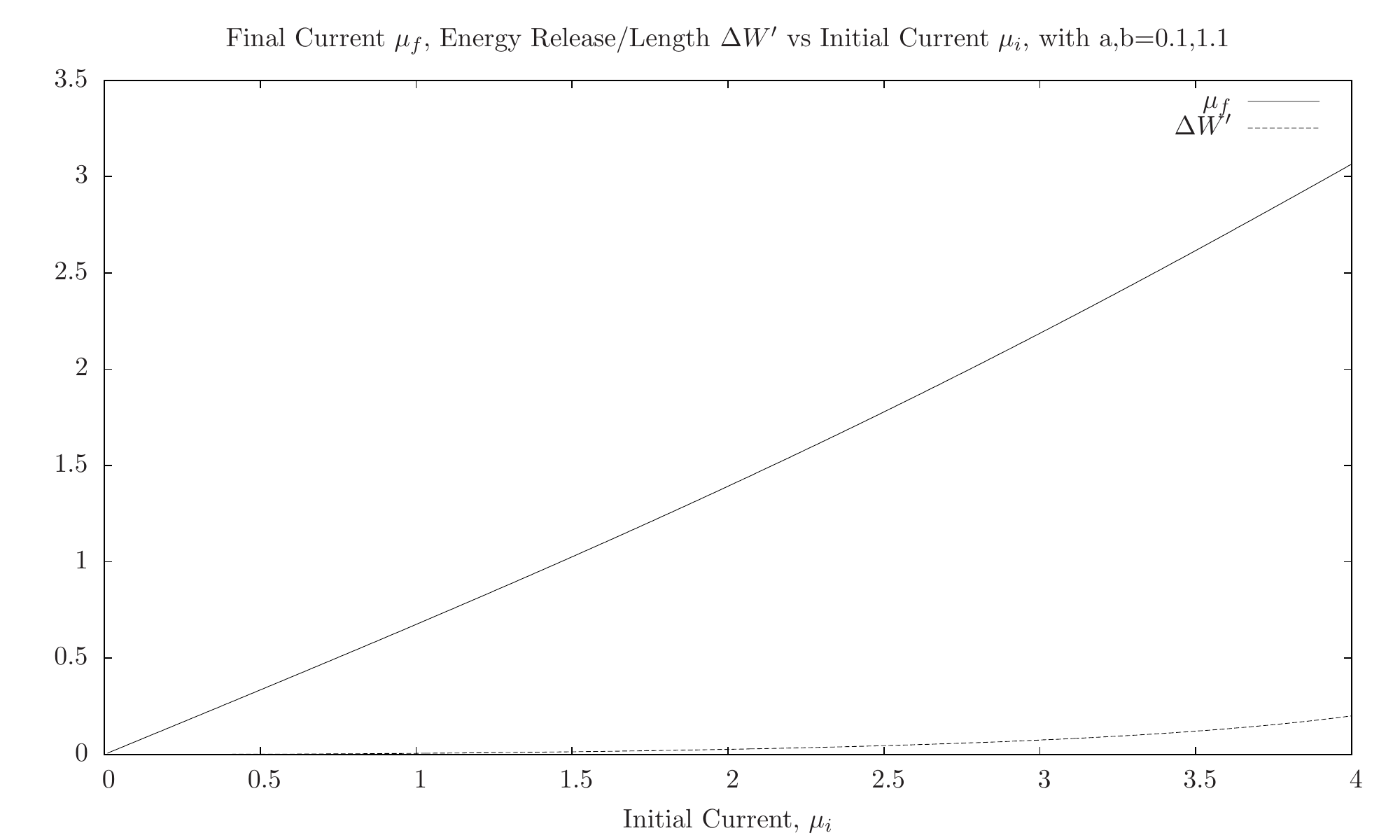}
\includegraphics[width=0.7\textwidth]{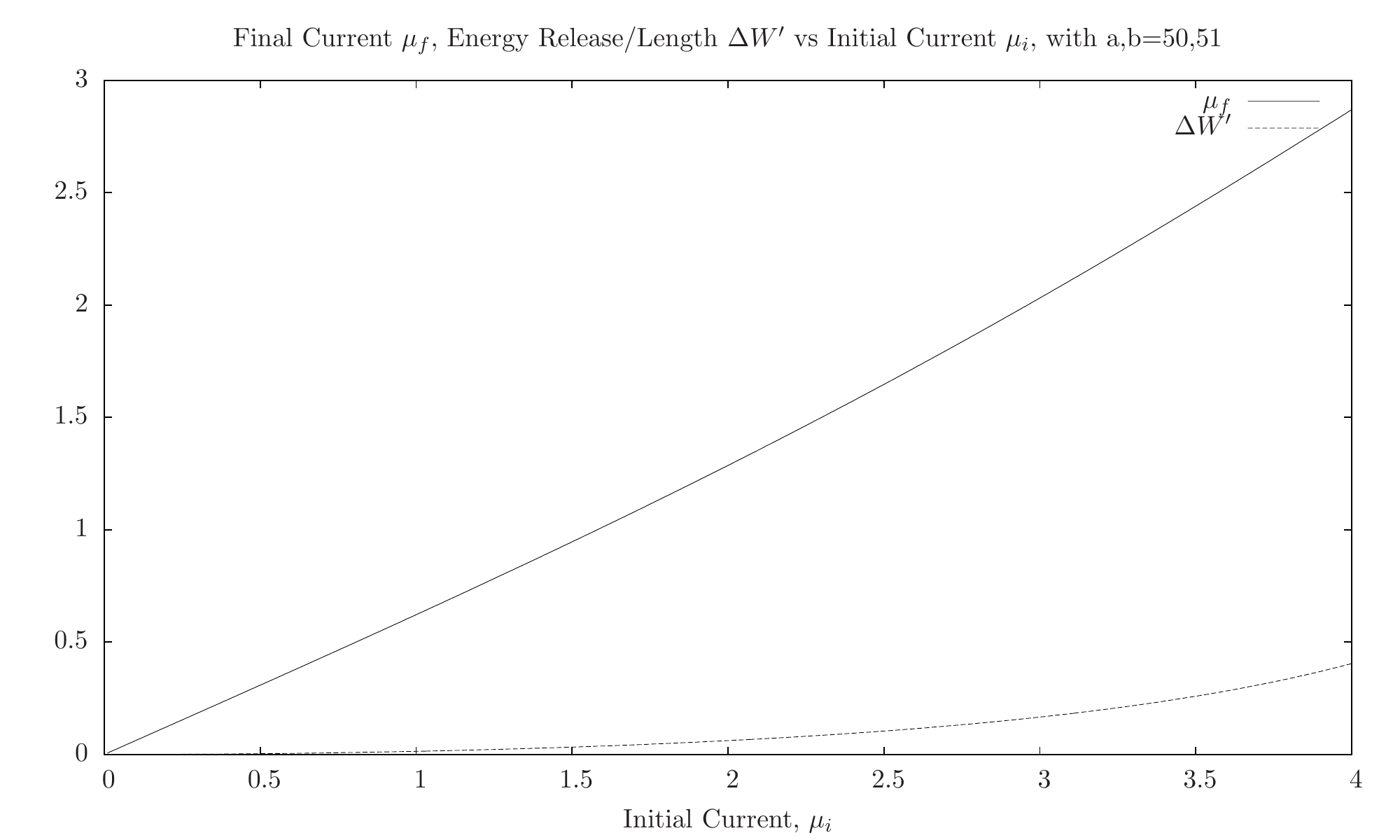}
\caption{Final $\mu =\mu_f$ (solid curve) and energy released, $\Delta W'$ (dashed curve), as a function of the initial value $\mu_i$ (dimensionless units), for the relaxation model in a cylindrical  configuration with (top)  $a = 0.1, b= 1.1$ and (bottom) $a = 50, b = 51$.}
\label{mu_dependence}
\end{figure}

Since the toroidal plasma current is $I_t=\mu\Phi_{t}$, Fig. \ref{mu_dependence}  may also be interpreted, re-scaling the axes by the appropriate fluxes, as showing the dependence of final plasma current on initial current in one flux rope. However, the initial  state also includes a current sheet, and in order to calculate the {\it total} initial plasma current, we must add the (negative) net current in this sheet, which may be simply obtained as $I_{t,cs}=-(2/\mu_{0})\int\limits{_a^b}B_{r}(r,0)dr$.
 
It can be shown analytically that for small $\mu_{i}$, $\Delta W \sim \mu_{i}^2$, $\mu_{f} \sim \mu_{i}$. Hence the final plasma current depends linearly on the initial current; whilst the energy release depends quadratically on $\mu_{i}$ and hence also on the initial poloidal field. This agrees with experimental findings in MAST and TS-3 \cite{ono12}. The dependence of  $\mu_{f}$ on $\mu_{i}$ shown in Figure 5 appears as a straight line, but the scaling is  in fact stronger than linear, which  becomes evident  for larger $\mu_{i}$ as there is a singularity at the first eigenvalue \cite{browning93}. Such large values of current are not relevant to spherical tokamaks, but may be found in solar coronal loops (see Section \ref{corona}).

As shown in Section \ref{simulations}, the magnetic energy released by reconnection may both heat the plasma and drive plasma flows - but the latter may also be dissipated to provide further heating. We obtain an estimate (strictly an upper bound) on the average temperature increase - here simply within single-fluid MHD - by assuming that the released magnetic energy is converted fully into thermal energy. Taking typical MAST parameters and equating the energy released to the gain in thermal energy, we estimate an average temperature increase of 100 eV, which is compatible with what is experimentally  observed  - of course,  the relaxation model cannot predict any spatial distribution of temperature. 
 
\section{Merging flux ropes in the solar corona}
\label{corona}
The simulations of merging flux ropes described in Section \ref{simulations} may be of some significance in understanding reconnection in the solar corona. The latter is, like MAST,  a low $\beta$ plasma, and whilst the Lundquist number in MAST (up to $10^7$) is much lower than the coronal value (around $10^{12-14}$), MAST comes closer to this highly-conducting regime than other laboratory reconnection experiments. Coronal reconnection is widely modelled using MHD, within which framework the classical "Sweet-Parker" model predicts reconnection rates which are far too slow to explain coronal heating or solar flares, suggesting that physics beyond single-fluid MHD may be required \cite{demoortel15}. Furthermore, the Sweet-Parker current sheet width in the solar corona is comparable with or smaller than the ion-skin depth, also indicating that Hall physics (at least) must be taken into account \cite{browning14}. Indeed, analysis of regimes of reconnection suggests that this simple comparison of length-scales may  under-estimate the importance of incorporating Hall physics \cite{ji11}. The features of reconnection dynamics  identified through  the Hall MHD simulations described in \cite{stanier13} and in Section \ref{simulations} above should thus be of some relevance to the solar corona. Resistive MHD simulations of merging flux ropes in the corona, brought together by attraction of like currents, have been been performed by  Kondrashov et al \cite{kondrashov99}; recently, a rather different scenario in which merger is triggered by kink instability in one flux rope has also been demonstrated \cite{tam15}.

In order to explain solar coronal heating, it is of primary importance to predict the dissipation of stored magnetic energy and, as first proposed by \cite{heyvaerts84}, relaxation theory thus provides a very useful framework. Browning et al \cite{browning86} used  an  approximate relaxation  model valid for fields close to potential, to predict heating rates in a set  of adjacent twisted flux tubes. The relaxation model of merging flux tubes \cite{browning14}  may thus be adapted to model coronal heating. Since coronal loops  have only weak curvature, and are not complete tori, the infinite-aspect-ratio model, in which the initial state  comprises  adjacent cylindrical twisted flux-ropes, is appropriate. The model has to be adapted for a loop of finite length $l$, line-tied to the photosphere at its ends: but it can be shown that the gauge-invariant definition of helicity used for an infinite-aspect-ratio system (periodic in  the axial direction) \cite{browning14} gives the relative helicity appropriate for a field with finite length and non-zero normal field at the boundaries. 

We note that the  energy release calculated in \cite{browning14} may be approximated by 
\begin{equation}
\Delta W \sim 0.1\mu_i^2 r^4 B^2 l/\mu_0,
\end{equation}
 for a loop of radius $r$, length $l$ and peak axial field $B$. Using the relation  $\mu \approx 2\phi/l$ where $\phi$ is the field-line twist-angle $\phi$ (valid near the loop axis) \cite{browning89}, and taking typical coronal values $B = 0.01\,$T, $l = 10\,$Mm, $r = 1\,$Mm, we find for rather weakly-twisted loops with $\phi = \pi$ an energy release of around $10^{19}$ J. Complete conversion of this energy into thermal energy gives a temperature increase of around $10^7$ K for a plasma of density $10^{15} m^{-3}$. This suggests that such flux tube mergers - if occurring sufficiently frequently - could make a significant contribution to coronal heating, and might be observable as microflares.

Twisted flux ropes in the corona may arise either through emergence from below the surface or by twisting through vortical photospheric motions. Both processes allow for the possibility that neighbouring twisted filaments may have twists in opposite directions (a situation not possible in the MAST experiment). We therefore consider the relaxation of two flux ropes with  $\mu= \pm \mu_i$, hence having the same fields but with twists in opposing directions  - this is the counter-helicity case, whereas the case previously discussed is co-helicity. The total helicity is thus zero, and  the relaxed state is therefore potential with  $\mu_f = 0$. Hence the energy release is significantly greater than for two flux tubes with the same twist, as shown in Figure 6. However, the initial configuration has no current sheet since the poloidal field is continuous across the interface $z=0$; thus, reconnection and merging is less likely in this configuration;  if it happens at all - perhaps due to an external trigger - it should proceed at a slower rate. On the other hand, evidence of such counter-helicity merging is indeed observed in some solar flares \cite{liu07}, and the large energy energy release in such events is consistent with our  predictions.

\begin{figure}
\label{oppositetwist}
\center
\includegraphics[width=0.7\textwidth]{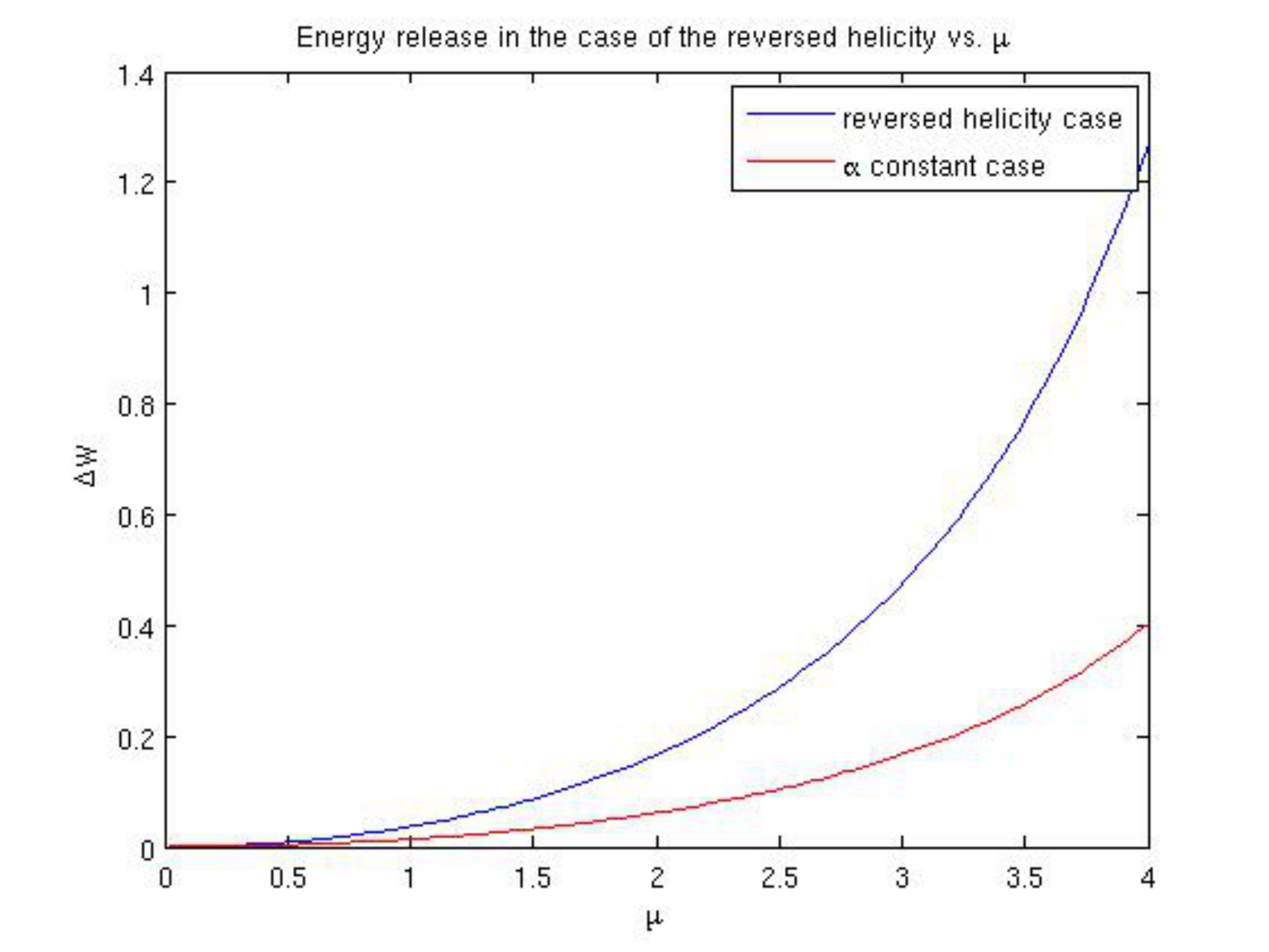}
\caption{Energy release as a function of the initial normalised current ($\mu_i$) for a pair of flux tubes with twist in the same direction (red) and opposite directions (blue). }
\end{figure}

According  to the ''flux tube tectonics'' scenario \cite{priest02}, the solar corona may contain many twisted filaments, which might be twisted in mixed directions. In reality, the $\mu$ values of such filaments may cover a range of values, but our analytical relaxation model allows us to model only the case of an array of filaments for  which the magnitudes of $\mu$ are all equal (although the signs may vary, allowing for the possibility of filaments twisted opposing senses). As an initial state, we assume a $n =N \times N$ square array of flux-ropes each with $\mu= \pm \mu_i$. First, consider the case of four tubes ($N=2$), so there are three distinct possibilities for the signs of $\mu_i$: all positive, three positive, one negative; two of each sign. (The relative positions of the different orientations is irrelevant to the energy release in the relaxation model, which is also unaffected by an overall change in sign of twist). As expected, the energy release is greatest in the case when equal numbers of positively and negatively twisted filaments are present, so that the total helicity vanishes and the minimum energy state is potential, see Figure 7.

\begin{figure}
\label{fourtubes}
\center
\includegraphics[width=0.7\textwidth]{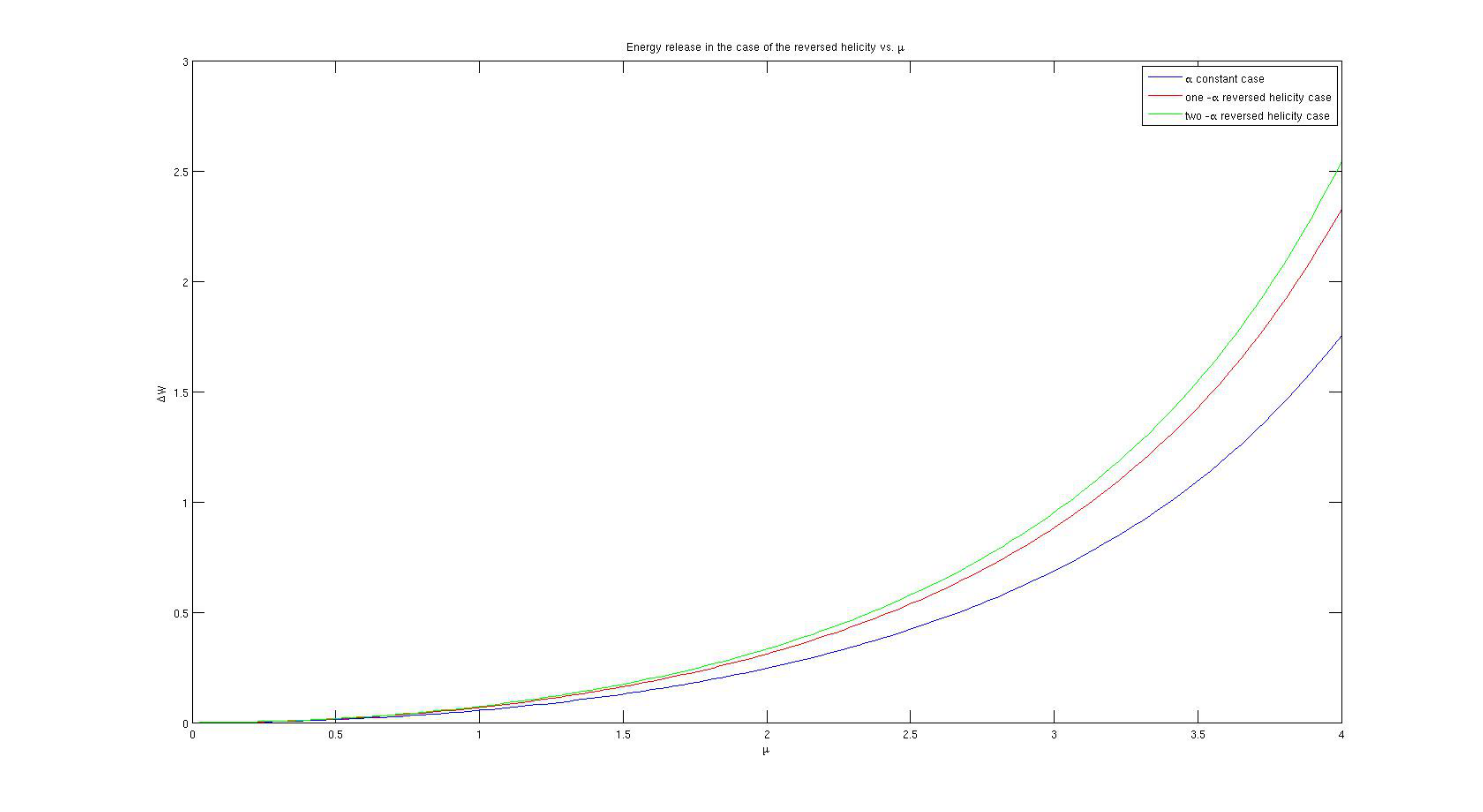}

\caption{Normalised energy release as a function of the initial normalised current ($\mu_i$) for an array of four flux tubes with twists: all in the same sense; three in one sense, one opposite; two in each sense.}

\end{figure}

We then investigate the effect of varying the number of initial filaments, $n$. The variation of energy release with initial $\mu$ is shown in Figure 8. Here, the magnetic flux and the total cross-section are kept constant, so that the width of each initial twisted filament is $a/N$. In all cases, the final relaxed state consists of a single twisted flux tube; the energy released increases with the number of flux tubes, due mainly to the increasing number of current sheets. The scaling of energy release with number of merging tubes, for a fixed $\mu_i$ is found to be  very close to linear $\Delta W \sim n^{1.07}$. This exemplifies relaxation as an inverse-cascade from small-scale to large-scale structure.

\begin{figure}
\label{nfluxtubes}
\center
\includegraphics[width=0.7\textwidth]{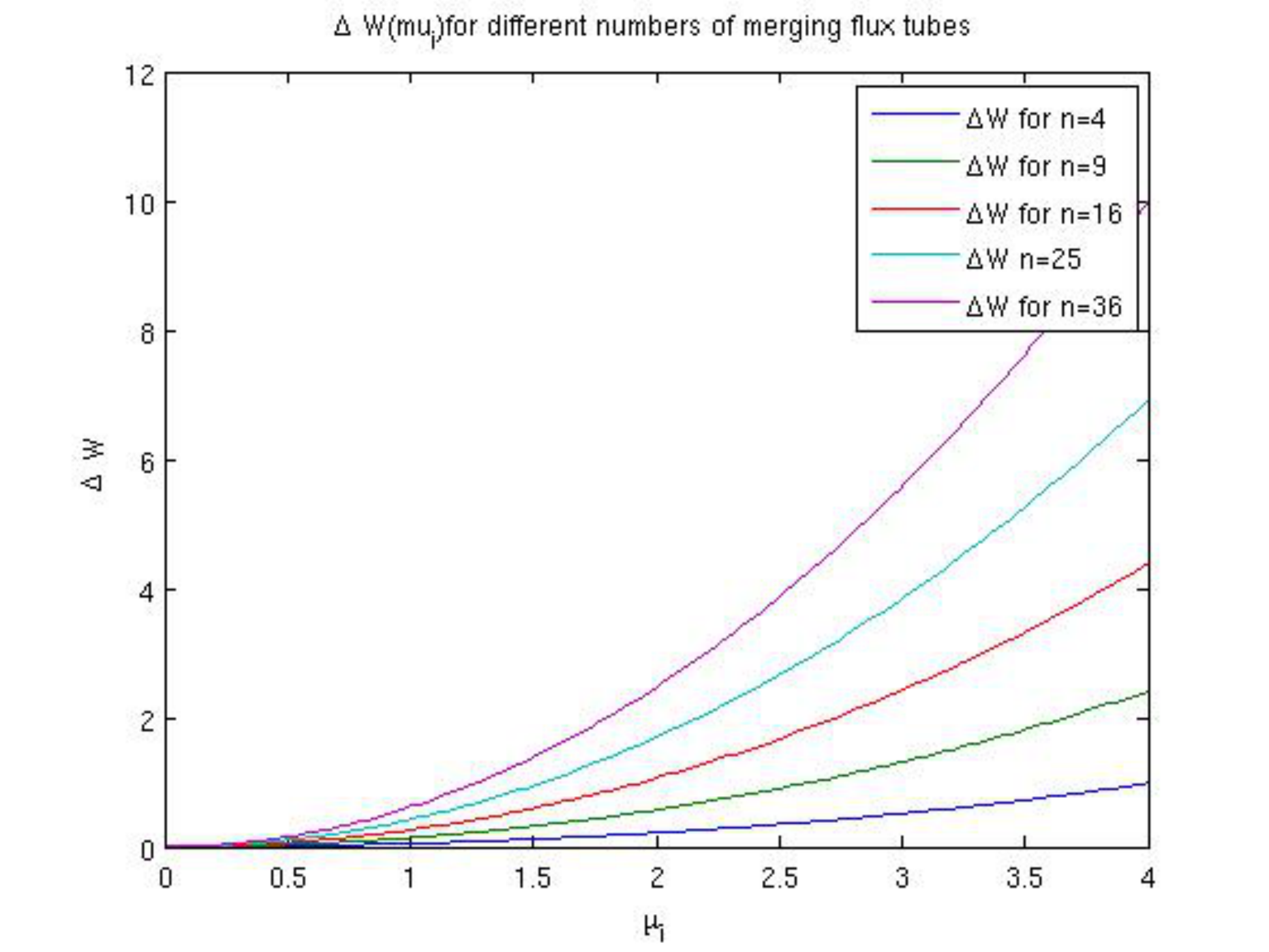}

\caption{Normalised energy release as a function of the initial normalised current ($\mu_i$) for different numbers of flux tubes twisted in the same sense, initially in a square array within a fixed cross-section: n = 4 (blue), 9 (green) , 16 (red), 25 (cyan), 36 (purple).  }

\end{figure}

\section{Conclusions}
We have described magnetic reconnection during the merging of magnetic flux ropes in the MAST spherical tokamak and in the solar corona - in both cases, reconnection leads to the formation of larger-scale magnetic structures and the dissipation of free magnetic energy. Simulations of the merging using 2D resistive MHD and Hall MHD have been performed, showing in detail how the reconnection rate and dynamics depend on the collisionality. The simulations predict the formation of a  spherical tokamak  with a single set of closed flux surfaces; the temporal evolution of the density profile matches experimental results. The inclusion of the Hall term in Ohm's law permits much faster reconnection. Whilst the ion skin depth is relatively much smaller, in comparison with global length scales,  in the solar corona, the Hall term is also likely to be significant there, and the simulations are indicative of processes which may be occurring in the corona. Simulations show that the magnetic reconnection is associated with significant heating of both ions and electrons. Future work will model reconnection in merging solar flux ropes, taking account of the Hall term.

An analytical model based on helicity-conserving relaxation to a minimum-energy state allows prediction of the final field configuration and the energy release, which may be easily applied to different configurations  and parameters. The average temperature increase can be calculated if it is assumed that all the dissipated magnetic energy is converted to thermal energy. We described  how a previous model assuming infinite aspect-ratio can be extended to finite aspect-ratio geometry, with particular relevance to spherical tokamaks. The infinite-aspect ratio model provides a good approximate model, but the energy release for a given initial ratio of current to field is approximately a factor of two lower in the finite aspect-ratio model for MAST parameters. 

The infinite-aspect-ratio relaxation model has been adapted to apply to coronal loops. Simple estimates of the temperature increase associated with the merging of two weakly-twisted coronal loops suggest this process could contribute significantly to coronal heating. In the coronal case, adjacent flux tubes may be twisted in the same or in opposite senses - the latter case leads to significantly larger energy release, but further simulations are needed in future to explore the conditions in which such flux tube merging can occur. A single coronal loop may consist of multiple twisted threads. Within a given volume, the energy release increases with the number of twisted threads, correlating with the number of current sheets.

\label{conclusions}

\section{Acknowledgements}
This work was funded by the UK STFC, the US DoE Experimental
Plasma Research program, the RCUK Energy
Programme under grant EP/I501045, and by  Euratom.  The
views and opinions expressed herein do not necessarily
reflect those of the European Commission.  Any opinion, findings, and conclusions or recommendations expressed in this material are those of the authors and do not necessarily reflect the views of the National Science Foundation.


\end{document}